\documentclass{aa}
\usepackage{graphicx}
\newcommand{\kms}{${\rm km \, s^{-1}}$}
\def\lesssim{\mathrel{\hbox{\rlap{\hbox{\lower4pt\hbox{$\sim$}}}\hbox{$<$}}}}
\def\gtrsim{\mathrel{\hbox{\rlap{\hbox{\lower4pt\hbox{$\sim$}}}\hbox{$>$}}}}
\def\RA#1#2#3#4{#1\std&#2\min&#3\dsec&#4} 	
\def\DEC#1#2#3{$#1$\grad&$#2$\bmin&$#3$\bsec}  	

\def\grad{\hbox{$^\circ$}}

\def\bmin{\hbox{$^\prime$}}

\def\bsec{\hbox{$^{\prime\prime}$}}

\def\std{\hbox{$^{\rm h}$}}

\def\min{\hbox{$^{\rm m}$}}

\def\dsec{\hbox{$.\!\!^{\rm s}$}}
\begin{document}
   \title{High resolution spectroscopy of bright subdwarf B
	        stars\thanks{Based on observations collected at the German-Spanish 
	Astronomical Center (DSAZ), Calar Alto, operated by the 
	Max-Planck-Institut f\"ur Astronomie Heidelberg jointly
	with the Spanish National Commission for Astronomy.}\fnmsep\thanks{Based
  on observations collected at the  
        European Southern Observatory at La Silla, Chile,
        ESO proposal No. 65.H-0341(A), 69.D-0016(A), 
        and 073.D-0495(A)}\fnmsep
    }

   \subtitle{I. Radial velocity variables}

   \author{H.~Edelmann\inst{1},
          U. Heber\inst{1}, M. Altmann\inst{2}, C. Karl\inst{1}, T. Lisker\inst{3}
          }

   \offprints{H. Edelmann, \email{edelmann@sternwarte.uni-erlangen.de}}

   \institute{Dr. Remeis-Sternwarte, Astronomisches Institut der Universit\"at~Erlangen-N\"urnberg,
              Sternwartstrasse~7, D-96049~Bamberg, Germany
         \and
             Departamento de Astronomia de la Universidad de Chile, Casilla 36D, Correo Central,
             Santiago, Chile
				 \and
             Astronomical Institute, University of Basel, Venusstrasse~7, CH-4102~Binningen, Switzerland
\\
             }

   \date{Received April 19, 2005; accepted ????? ??, 2005}

   \abstract{Radial velocity curves for 15 bright subdwarf B 
binary systems have been measured using high precision radial velocity measurements from 
high S/N optical high-resolution spectra. 
In addition, two bright sdB stars are discovered to be radial velocity variable
but the period could not yet be determined. The companions for all systems are unseen. 
The periods range from about 0.18 days up to more than 
ten days.
The radial velocity semi amplitudes are found to lie between 15 
and 130 \kms.
Using the mass functions, the masses of the unseen companions 
have been constrained to lower limits  
of 0.03 up to 0.55 ${\rm M}_{\sun}$, and most probable values of 
0.03 up to 0.81 ${\rm M}_{\sun}$. The invisible companions
for three of our program stars are undoubtedly white dwarfs.
In the other cases they could be either white dwarfs or main sequence stars.
For two stars the secondaries could possibly be brown dwarfs.
%
%
As expected, the orbits are circular for most of the systems.
However, for one third of the program stars we find slightly eccentric orbits 
with small eccentricities of $\epsilon\approx 0.02$--0.06.
This is the first time that non-circular orbits have been found in sdB
binaries. 
No correlation with the orbital period can be found.
%
   }

   \authorrunning{Edelmann et al.}
   \maketitle
%

\section{Introduction}
While the evolutionary status of subdwarf B (sdB) stars as extreme horizontal 
branch stars is well established (e.g. Heber et al. \cite{Heb84}, 
Heber \cite{Heb86}, Saffer et al. \cite{Saf94}, 
Maxted et al. \cite{Max01}, Edelmann et al. \cite{Ede03a}, and others), 
their origin has been under discussion for years. 
Indications that many sdB stars are members of binary 
systems suggest that binary interaction is important 
for their evolution.
In addition to composite spectrum binaries 
discovered e.g. by Ferguson, Green \& Liebert (\cite{Fer84}), 
Allard et al. (\cite{All94}), Theissen et al. (\cite{The93}, \cite{The95}), 
Jeffery \& Pollacco (\cite{Jef98}), Ulla \& Thejll (\cite{Ull98}), 
and others, several single-lined binary sdB stars have been identified 
from variable Doppler line shifts resulting from orbital motion 
(e.g. Green, Liebert \& Saffer \cite{Gre01}, Maxted et al. \cite{Max01}).
On the other hand, there are also many subdwarf B stars showing no 
indication of any companion. However, this can also be explained by
close binary interaction.
From a theoretical point of view, Han et al. (\cite{Han02}, \cite{Han03}) 
elucidated in detail three channels that can produce sdB stars from 
close binary systems: 
(i) The common envelope ejection channel with one or two common envelope 
phases.
The first phase results in a sdB and a main sequence star 
with periods between 0.05 and 40 days, the second phase results in a binary 
with a 
white dwarf companion and a wider range of orbital periods.
(ii) The stable Roche lobe overflow channel: results in a binary
with a main sequence secondary and a wide range of orbital periods 
of 0.5 days up to 2000 days. 
(iii) The merger channel: two helium white dwarfs merge into one single 
sdB star. 

A recent comparison of a large observational sdB sample with these models 
demonstrated that in principle, all sdB stars could be produced by close binary 
evolution when adopting the three above channels (Lisker et al. \cite{Lis05}).
However, there is disagreement between the observational and theoretical samples 
concerning the fractional/differential population of the Extended Horizontal Branch 
(EHB) by sdB stars, which still leaves room for the possibility of a single-star
evolutionary channel. Therefore, more detailed radial velocity studies of
sdB stars 
are crucial for a reliable
evolution of current sdB formation theories.

In a series of papers we will present detailed analyses 
of a sample of more than five dozen of bright ($B\la13$~mag)
hot subluminous stars 
selected from the {\it Catalogue of Spectroscopically Identified 
Hot Subdwarf Stars}
(Kilkenny, Heber \& Drilling \cite{KHD88}, {\O}stensen \cite{Ost04}). 
These bright stars are ideally suited for detailed
studies, but apparently have been largely overlooked previously.
We want to 
derive the binary frequency, atmospheric parameters
and population characteristics (paper II), as well as the
metal abundances, isotopic anomalies and rotation velocities  
(paper III; some preliminary results are given by
Edelmann, Heber \& Napiwotzki \cite{Ede01}).

In this paper (paper I) we present the results of 17 bright, 
short period radial velocity (RV)
variable sdB systems amongst all analyzed bright subdwarf B star.
For 15 of them we could determine their 
orbital parameters with high accuracy.
Preliminary results for some of them have already been reported by
Edelmann et al. (\cite{Ede01}, \cite{Ede03b}, \cite{Ede04}).
Section~2 describes the observations and data reduction procedures. 
Section~3 is dedicated to the 
determination of the radial velocity curves, orbital parameters and the
resulting most probable nature of the unseen companions.
Finally, in Section~4 we discuss the results of our 
analysis and highlight the likely 
eccentric orbits for five program stars.
We summarize and conclude the paper in Section~5. 
%
\section{Observations}
\subsection{Program stars}
\label{program_stars}
%
All radial velocity variable sdB stars presented in this paper 
(see Table \ref{coordinates}
for the list of all stars including coordinates and B magnitudes)
have been selected from a sample of more than five dozen bright 
blue stars, 
in which we searched for RV variations 
(see forthcoming paper II).
%
%

Our search has already led to the discovery of a spectacular binary 
(\object{HD~188112}, Heber et al. \cite{Heb03})
consisting of a sdB star of too low mass ($M=0.24~{\rm M}_{\sun}$)
to sustain helium burning, and a massive compact companion ($M>0.7~{\rm M}_{\sun}$).
We obtained five new spectra for \object{HD~188112} to improve the orbital parameters.
%
\begin{table}[h]
\caption[Coordinates and B magnitudes for all program sdB stars ]
{Coordinates and B magnitudes for our program stars.
Detailed information concerning the observing runs is given in Table \ref{observations_hires}.} 
\label{coordinates}
\begin{tabular}{
									 l																															
									 @{\hspace{1mm}}l@{\hspace{-0.1mm}}l@{\hspace{-0.1mm}}l@{\hspace{-0.1mm}}l   	
									 @{\hspace{2mm}}l@{\hspace{-0.1mm}}l@{\hspace{-0.1mm}}c												
                   @{\hspace{2mm}}l																															
									 @{\hspace{-2mm}}l   																													
                   }\hline																											  
& \multicolumn{4}{l}{\hspace{-2mm}$\alpha$} 
& \multicolumn{3}{l}{\hspace{-1mm}$\delta$} 
& \multicolumn{1}{l}{\hspace{-2mm}$B$} 
& run 
\\ 
star 
& \multicolumn{4}{l}{\hspace{-2mm}(2000)} 
& \multicolumn{3}{l}{\hspace{-1mm}(2000)} 
& \multicolumn{1}{c}{\hspace{-4mm}(mag)} 
& \#
\\ \hline
\object{Ton~S~135}              &  \RA{00}{03}{22}{1} & \DEC{-23}{38}{58} & 13.0 & 7       \\
\object{PG~0001+275}            &  \RA{00}{03}{55}{7} & \DEC{+27}{48}{37} & 12.6 & 1,4    \\
\object{Ton~S~183}              &  \RA{01}{01}{17}{6} & \DEC{-33}{42}{45} & 12.5 & 7       \\
\object{PG~0133+114}            &  \RA{01}{36}{26}{1} & \DEC{+11}{39}{33} & 12.1 & 2,4    \\
\object{CD$-24\degr$~731}       &  \RA{01}{43}{48}{4} & \DEC{-24}{05}{10} & 11.6 & 3,7    \\
\object{HE~0230$-$4323}         &  \RA{02}{32}{54}{6} & \DEC{-43}{10}{28} & 13.5 & 7       \\
\object{BPS~CS~22169$-$0001}    &  \RA{03}{56}{23}{3} & \DEC{-15}{09}{20} & 12.6 & 7,8    \\
\object{CPD$-64\degr$~481}      &  \RA{05}{47}{59}{3} & \DEC{-64}{23}{03} & 11.1 & 7,8    \\
\object{PG~1232$-$136}           &  \RA{12}{35}{18}{9} & \DEC{-13}{55}{09} & 13.1 & 8       \\
\object{[CW83]~1419$-$09}           &  \RA{14}{22}{40}{3} & \DEC{-09}{17}{20} & 11.7 & 8       \\
\object{[CW83]~1735+22}            &  \RA{17}{37}{26}{5} & \DEC{+22}{08}{58} & 11.6 & 1,4    \\
\object{HD~171858}              &  \RA{18}{37}{56}{7} & \DEC{-23}{11}{35} & 10.4 & 3,5,6 \\
\object{HD~188112}              &  \RA{19}{54}{31}{4} & \DEC{-28}{20}{21} & 10.0 & 3,5-7  \\
\object{JL~82}                  &  \RA{21}{36}{01}{2} & \DEC{-72}{48}{27} & 12.2 & 7       \\
\object{PB~7352}                &  \RA{22}{55}{43}{2} & \DEC{-06}{59}{39} & 12.1 & 3,5-7  \\
\object{LB~1516}                &  \RA{23}{01}{56}{0} & \DEC{-48}{03}{46} & 12.7 & 3,5,7 \\
\object{PHL~457}                &  \RA{23}{19}{24}{5} & \DEC{-08}{52}{37} & 12.7 & 7       \\
\hline
\end{tabular}\\
\end{table}
\begin{table}[h]
\caption[Echelle high-resolution observations]{Summary of the observations.}
\label{observations_hires}
\centering
\begin{tabular}{llll}\\\hline
run  & date                 & observatory  					& observers \\
\#			& (start of nights)		 & 						       		&						\\\hline
1       & 1999 Jul 19-23       & DSAZ    & E/P \\
2       & 2000 Jan 28 - Feb 01 & DSAZ    & E/K \\
3       & 2000 Sep 06-09       & ESO     & E\\
4       & 2001 Aug 27-31       & DSAZ    & E/K \\
5       & 2002 Aug 07-10       & ESO     & L\\
6       & 2002 Aug 14-21, Nov 26 & DSAZ    & sm\\
7       & 2004 Oct 31 - Sep 04 & ESO     & A\\
8       & 2005 Feb 23 - Mar 01 & ESO     & A\\
\hline\\
\end{tabular}
\parbox[t]{8cm}{observers:
    A~=~Altmann~M.;
    E~=~Edelmann~H.;
    K~=~Karl~C.;
		L~=~Lisker~T.;
    P~=~Pfeiffer~M.;
    sm~=~service-mode
}
\end{table}
\subsection{Observations and data reduction}
Optical echelle spectra with high S/N were obtained at two observatories:

80 spectra  were obtained at the 
German-Spanish Astronomical Center (DSAZ) on Calar Alto, Spain,
with the 2.2~m telescope equipped with the 
Fiber-Optics Cassegrain Echelle Spectrograph (FOCES). 
We used the Tektronic CCD Chip (1024$\times$1024 pixel) with a pixel size
of 24~$\mu$m, the 200$~\mu$m entrance aperture, and a slit width of 2 arc secs,
resulting in a nominal resolution of $\lambda/\Delta\lambda = 30\,000$.
The DSAZ spectra cover the wavelengths from 3\,900~\AA\ to 6\,900~\AA. 
%
%
The spectra were reduced as described in
Pfeiffer et al. (\cite{Pfe98}) using the IDL macros developed
by the Munich Group.

191 spectra were obtained at the European Southern Observatory (ESO)
on La Silla, Chile, with 
the  Fiber-fed Extended Range Optical Spectrograph (FEROS) mounted 
until October 2002 
on 
the 1.52~m ESO telescope, and afterwards 
on 
the 2.2~m telescope.
We used the standard setup (EEV CCD Chip with 2048$\times$4096 pixel, 
pixel size of 15~$\mu$m, entrance aperture of 2.7 arc secs)
with a nominal resolution of $\lambda/\Delta\lambda = 48\,000$.
The ESO spectra cover the wavelengths from 3\,600\AA\ to 8\,900\AA. 
The spectra were reduced using the on-line data reduction provided
at ESO 
(Pritchard \cite{Pri04})
applying the ESO-MIDAS program package.
%

For the summary of the observing runs performed for our
program stars  see Table \ref{observations_hires}.
\section{Analysis}
\subsection{Radial velocity measurements}
\label{radial_velocities}
\begin{figure}
\vspace{3.5cm}
\includegraphics{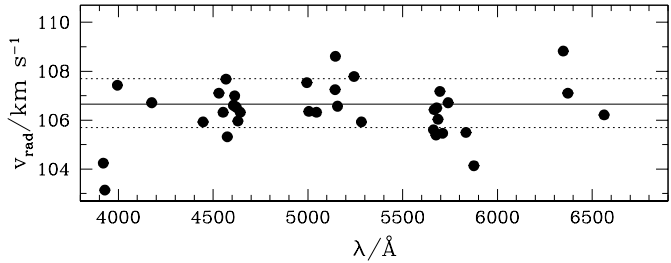}
	\caption[RVV sample plot]{
	Radial velocity determination for one spectrum of CPD$-64\degr$~481.
	The dots denote the radial velocities determined for all measured
	lines. 
	Note the very good consistency irrespective of the wavelength.
	The straight line indicates the mean value, and the dashed lines
	the $1\sigma$ error limit. 
	\label{vrad_plot}
}
\end{figure}

\begin{table*}[h]
\caption[RV measurements for all program stars]
{RV measurements for all program stars.} 
\label{rv_values}
\centering
\scriptsize
\include{rv_values}
\normalsize
\end{table*}
The radial velocities are determined by calculating the shifts of the 
measured wavelengths of 
\ion{He}{i}\,5876~\AA, and 
all clearly identified metal lines (mostly \ion{Si}{iii} triplet at 4553~\AA, 
4568~\AA, and 4575~\AA, \ion{N}{ii} at 4666~\AA, 5667~\AA, and 5680~\AA, and 
\ion{Si}{ii} at 6347~\AA, and 6371~\AA) 
to laboratory wavelengths.
We used the ESO-MIDAS software package to fit Gaussian curves to the absorption
lines in order to determine their central wavelengths.
After the measurement, the DSAZ values were corrected to heliocentric values.
Due to the automatic correction for the earth's  movement of the ESO spectra
during the on-line reduction process,
the measured wavelengths correspond to heliocentric values.

The errors for the given RV values which are derived from the measurements
of single lines are unrealistically small (typically: $\sim$0.1~\kms).
The syste\-ma\-ti\-c errors that arise from the observations (placement of the
stars disc on the slit, S/N), and
from the data reduction (e.g. wavelength calibration)
are dominant. 
To estimate the real errors we plotted for all single spectra the derived
RVs for all measured absorption lines versus the corresponding wavelength 
positions. This is exemplarily shown for one
spectrum  in Fig. \ref{vrad_plot}. 
No wavelength dependent trend for the obtained RV 
values could be found for any of our observations.
From each plot the mean RV value together with a "realistic" $1\sigma$
error limit (which is in most cases about $1-2$~\kms) can be determined. 

The resulting RV's for all program stars together with the individual errors 
are listed in Table 3. 
%
%
%
\subsection{Orbital parameters}
\label{period_determination}
\begin{figure*}
\vspace{12.5cm}
\includegraphics{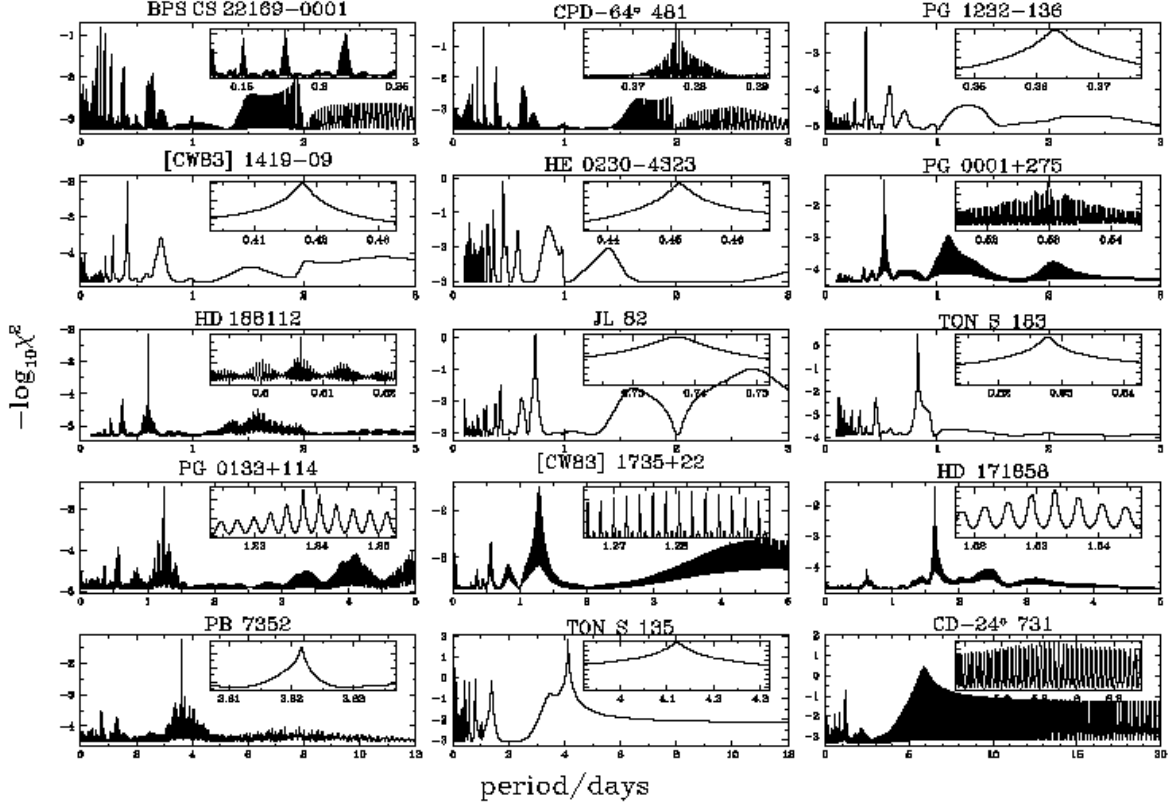}
	\caption[power spectra]
   {Power spectra of the measurements, together with details of the region around the main peak
    (ordered by increasing periods: from  top left to bottom right).
  	\label{plot_phase_power_2}
}
\end{figure*}
%
The period search was carried out by 
means of a periodogram analysis based on the Singular Value 
Decomposition (SVD) method. A sine-shaped RV curve is fitted to the 
observations for a multitude of phases which are calculated as a function 
of period (see Napiwotzki et al. \cite{Nap01}). 
The difference between the observed radial velocities and the 
best fitting theoretical RV curve for each phase set is evaluated in terms 
of the logarithm of the sum of squared residuals ($\chi^2$) as a function 
of period, 
yielding the power spectrum of the data set which allows to determine the 
most probable period of variability (see e.g. Lorenz, Mayer \& Drechsel
 \cite{Lor98}). 

\begin{figure*}
\vspace{16.5cm}
\includegraphics{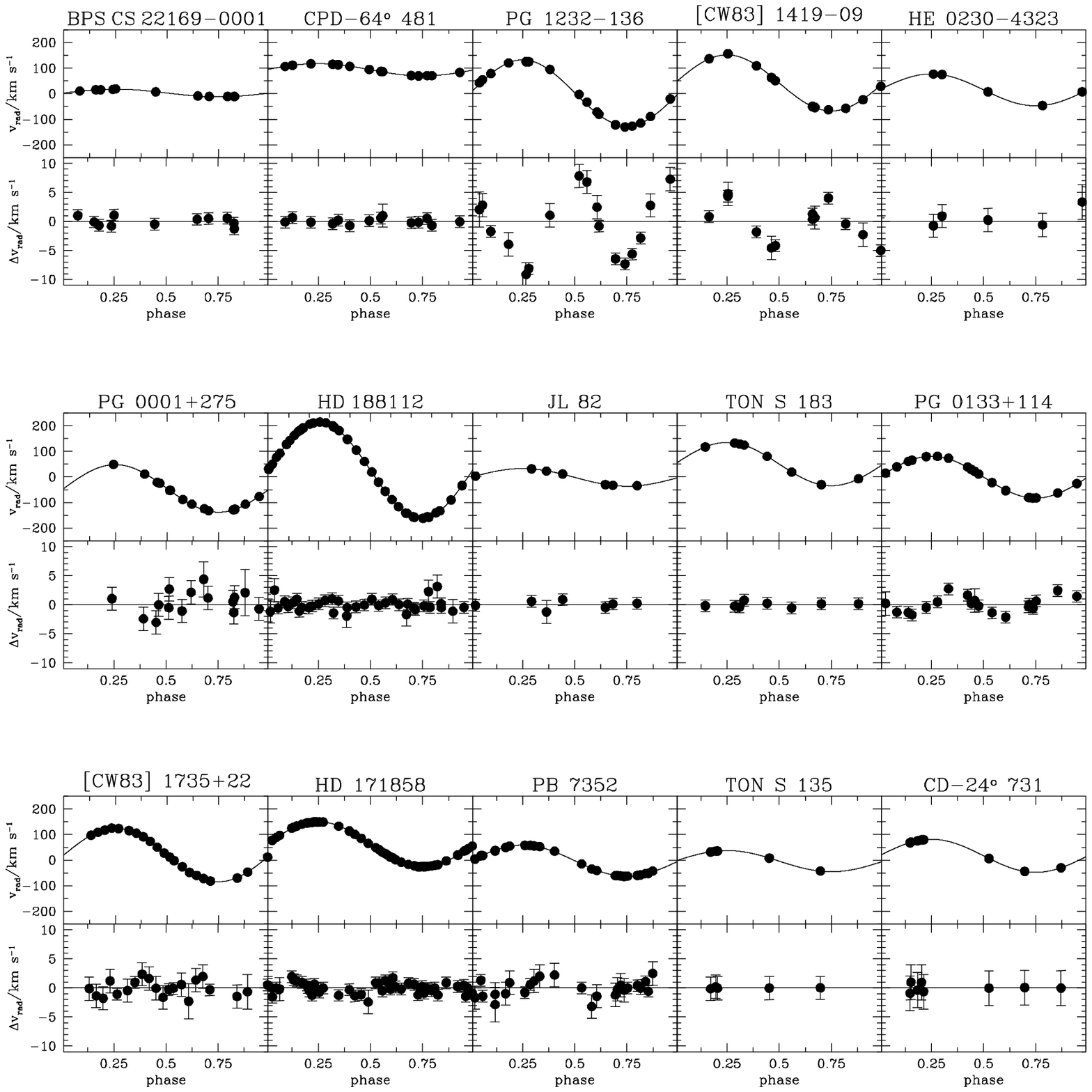}
	\caption[curves and residuals]
 {Measured radial velocities as a function of orbital phase and fitted sine curves, together with
the residuals ($\Delta v_{\rm rad}=v_{\rm obs} - v_{\rm sine}$)
to the sine fits in\-cluding error bars for all program stars
(ordered by increasing periods: from  top left to bottom right).
 	\label{plot_phase_power_1}
}
\end{figure*}

For twelve (70\%) of our program stars, the determined periods are 
unequivocal. 
But there are some cases, for which aliases exist which are less
likely but cannot completely be ruled out
(cf. Fig. \ref{plot_phase_power_2}).
\begin{itemize}
\item{\object{BPS~CS~22169$-$0001}:}
Beside the most prominent period of 0.1780(3) days
(number in parentheses give the uncertainty of the last given digit)
one alias exist at 0.2170(5) days.
\item{\object{CPD$-64\degr$~481}:}
Two periods of 0.276992(5) days and  0.277433(5) days, which corresponds 
to a time difference of 38~s ($\cor 0.0016 \times P $),
result from our analysis.
\item{\object{PG~0133+114}:}
An alias of $\Delta P=+53$~s exist (corresponding to an error for 
the period of 0.05\%).  
\item{\object{[CW83]~1735+22}:}
Many alias periods of $\Delta P\approx 3$~m ($\cor 0.0016 \times P$) exist. 
\item{\object{CD$-24\degr$~731}:}
There are many alias periods of $\Delta P\approx 34$~m ($\cor 0.004 \times P$).
\end{itemize}
However, for our purposes, a "perfect"  determination of the period is 
desirable but not necessary (see below).
Fig. \ref{plot_phase_power_2}
shows the resulting power spectra and 
Fig. \ref{plot_phase_power_1} shows the best fit RV curves for 15 stars.

\begin{figure}
\vspace{8.3cm}
\includegraphics{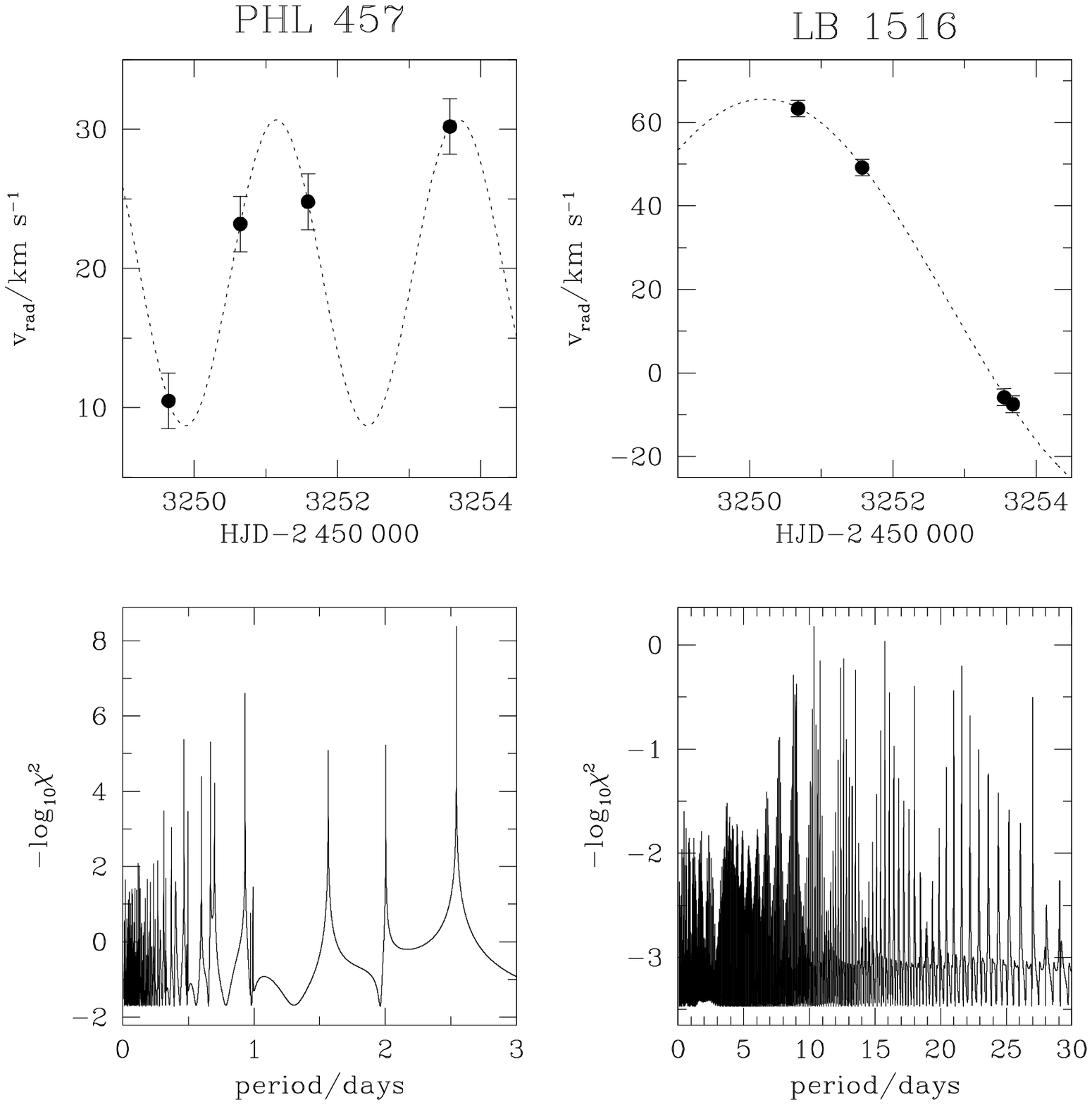}
	\caption[unsolved RV measurements]{
  Radial velocity measurements and power spectra (lower panel)
  for two radial velocity variable program stars for which no reliable or unique solution
  could be obtained yet. 
  Additionally are plotted the radial velocity curves (dotted lines)
  for the most prominent period.  
	Note that for LB~1516 only the RV values from the latest
	observing run (ESO Sep. 2004) are plotted to visualize the variability. 
	For the power spectrum of LB~1516 four additional points are included
  (cf. Table 3). 
	\label{rv_unsolved}
}
\end{figure}

Unfortunately, 
we could not determine unambiguous periods for two stars
(see Fig. \ref{rv_unsolved}).
\begin{itemize}
\item{\object{PHL~457}:}
Prominent periods are
2.54, 0.93, 0.47, 0.70, 2.00, and 1.57 days (ordered by power).
The number of measurements (only four) is too low to allow a reliable fit to 
be made.
At least, periods larger than three days cannot match the observations.
\item{\object{LB~1516}:} 
A large number of aliases are present. 
The most likely period is 10.36 days. 
Periods shorter than 9 days or larger
than 27 days can be excluded.
\end{itemize}
Table \ref{orbital_parameters} summarizes the orbital parameters (period,
ephemery, systemic velocity, and semi amplitude) for all analyzed stars.
%
%
%
%
%
\subsection{Eccentric orbits?}
\begin{figure*}
\vspace{7.5cm}
\includegraphics{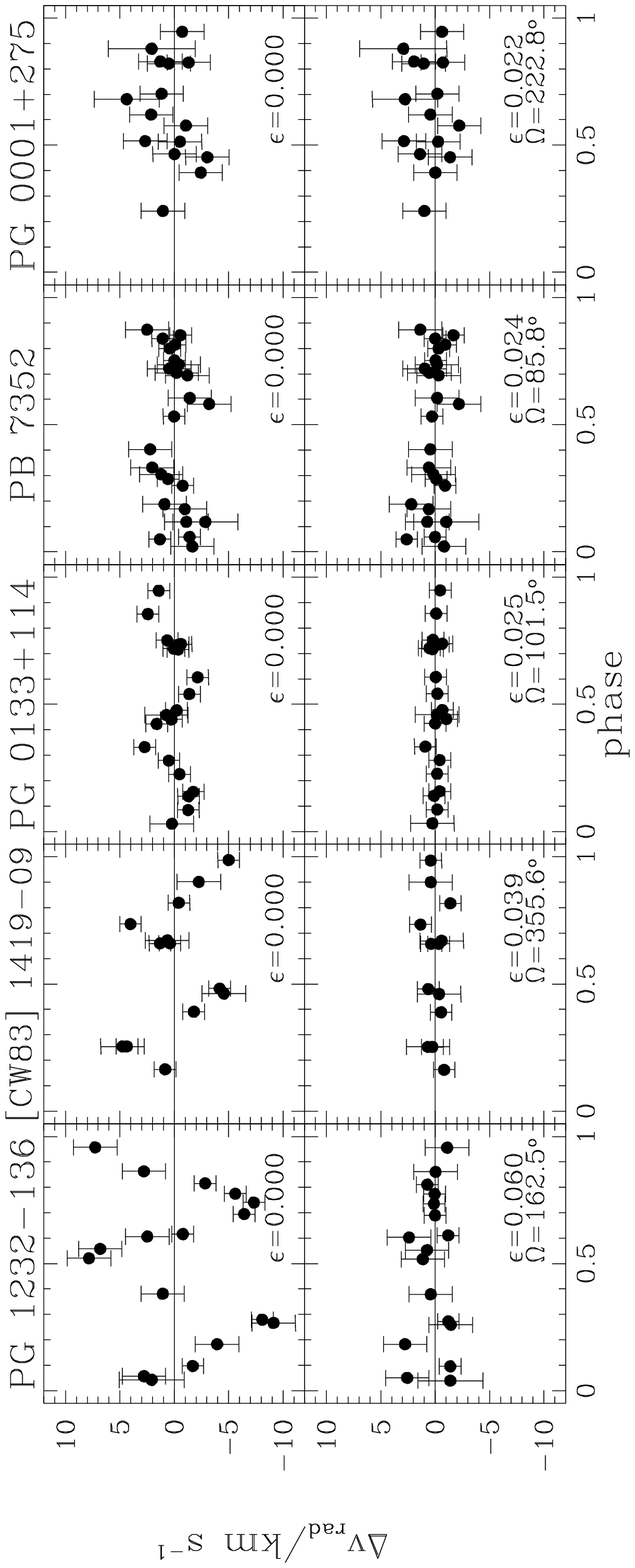}
	\caption[ecc for PG~0133+114]{
  Residuals to the fits including error bars for PG~1232$-$136, [CW83]~1419$-$09, PG~0133+114, 
  PB~7352, and 
	PG~0001+275 for two different eccentricities. 
	Upper part: Fits using vanishing eccentricities, Lower part:
	Best matching fits.
 	\label{plot_ecc}
}
\end{figure*}
Our high-precision measurements allow us to search for  
possible deviations from the adopted sine curve.

The residuals to the sine fits plotted in Fig. \ref{plot_phase_power_1}
indicate that the RV values for most of our program stars are well reproduced 
by sinusoidal curves (eccentricity $\epsilon<0.02$, i.e. our detection limit). 
This means that the orbits for
the majority of our analyzed stars are most likely circular.
However, for three stars (\object{PG~1232$-$136}, \object{[CW83]~1419$-$09}, and \object{PG~0133+114}), 
and less significantly also for \object{PB~7352} and \object{PG~0001+275}, 
periodic deviations of the residuals are noticed.
Triple systems can be ruled out on a very high confidence level,
because for all five systems the period of the remaining residuals 
are exactly one half of the binary orbital period, respectively. 
Observational or analytical effects are also not very likely. 
Especially for the most significant cases \object{PG~1232$-$136} and \object{[CW83]~1419$-$09},
all RV measurements are from one observing run. 
For both stars, during all nights the setup has not been changed, nor
has the data reduction, or analysis method. 
%
\object{PG~0133+114} also clearly shows 
periodic deviations of the residuals.
Its observations were carried out during different observing runs 
but at the same observatory (cf. Tables \ref{coordinates}, 
\ref{observations_hires}, and 3). 
For comparison, \object{[CW83]~1735+22} does not show any sign of
a periodic deviation of the residuals at all, although it
has almost the same period as
\object{PG~0133+114} and was observed mainly during the same observing
runs as \object{PG~0133+114} and \object{PG~0001+275}.
%

%
\begin{table*}[h]
\caption[Orbital parameters]
{Orbital parameters for our program stars, ordered by increasing periods from 
the top to the bottom.
Given are the periods $P$, 
the ephemeris for the time $T_0$ defined as the conjunction time 
at which the stars moves from the blue side to the red side of the RV curve,
the system velocities $\gamma_0$, the RV semi-amplitudes $K$,
the mass functions $f_m$, 
the masses of the system components ($M_{\rm comp.}^{i=90^\circ}$
and $M_{\rm comp.}^{i=52^\circ}$) assuming the canonical mass of
$M=0.5~{\rm M}_{\sun}$ for the sdB primary,
and the nature of the unseen companions due to their determined
masses (bd = brown dwarf, ms = main sequence star, wd = white dwarf).
All numbers in parentheses give the uncertainty of the last given digit.}
\label{orbital_parameters}
\centering
\begin{tabular}{lllr@{$\,\pm\,$}lr@{$\,\pm\,$}lllll}\hline
\multicolumn{1}{l}{star}          
& \multicolumn{1}{l}{\hspace*{2mm}$P$}                                
& HJD($T_0$)
& \multicolumn{2}{l}{\hspace*{2mm}$\gamma_0$}
& \multicolumn{2}{l}{\hspace*{2mm}$K$} 
& \multicolumn{1}{l}{$f_m$}
& $M_{\rm comp.}^{i=90^\circ}$\hspace*{-5mm}
& $M_{\rm comp.}^{i=52^\circ}$\hspace*{-5mm}
& nature
\\
\multicolumn{1}{c}{}
& \multicolumn{1}{l}{\hspace*{2mm}days}
& $-2\,450\,000$
& \multicolumn{2}{l}{\hspace*{2mm}[\kms]} 
& \multicolumn{2}{l}{\hspace*{2mm}[\kms]} 
& \multicolumn{1}{l}{[M$_{\sun}$]}
& \multicolumn{1}{c}{[M$_{\sun}$]}
& \multicolumn{1}{c}{[M$_{\sun}$]}
& comp.
\\
\hline
\object{BPS\,CS\,22169-0001} \hspace*{-5mm} & \, 0.1780(3)     & 3423.614(5)      &    2.8 & 0.3  &  14.9 & 0.4 & 0.00006(2)\hspace*{-10mm} & 0.03 & 0.03  & bd/ms/wd    \\        
\object{CPD$-64\degr$~481} & \, 0.2772(5)     & 3249.969(5)      &   94.1 & 0.3  &  23.8 & 0.4 & 0.00038(3) & 0.05 & 0.06 & bd/ms/wd    \\
\object{PG~1232$-$136}     & \, 0.3630(3)     & 3423.478(5)      &    4.1 & 0.3  & 129.6 & 0.4 &  0.0818(8) & 0.41 & 0.58 & ms/wd  \\    
\object{[CW83]~1419$-$09}     & \, 0.4178(2)     & 3424.842(5)      &   42.3 & 0.3  & 109.6 & 0.4 &  0.0567(6) & 0.34 & 0.49 & ms/wd \\     
\object{HE~0230$-$4323}    & \, 0.4515(2)     & 2151.573(5)      &   16.6 & 1.0  &  62.4 & 1.6 & 0.011(2)  & 0.17 & 0.22 & ms/wd    \\
\object{PG~0001+275}       & \, 0.529842(5)   & 2152.20358(5) \hspace*{-10mm}   & $-$44.7 & 0.5  &  92.8 & 0.7 & 0.043(1)  & 0.29 & 0.41 & ms/wd \\
\object{HD~188112}         & \, 0.6065812(5)\hspace*{-5mm}  & 2151.937898(5)   &   26.7 & 0.2  & 188.4 & 0.2 & 0.420(1)$^*$ & 0.73 & 1.22 & wd$^{**}$ \\
\object{JL~82}             & \, 0.7371(5)     & 2151.332(5)      &  $-$1.6 & 0.8  &  34.6 & 1.0 & 0.0031(3)  & 0.10 & 0.13 & ms/wd    \\
\object{Ton~S~183}         & \, 0.8277(2)     & 2151.586(5)      &   50.5 & 0.8  &  84.8 & 1.0 & 0.052(2)  & 0.32 & 0.45 & ms/wd \\
\object{PG~0133+114}       & \, 1.23787(3)    & 2151.2667(5)     &  $-$0.3 & 0.2  &  82.0 & 0.3 & 0.0712(2)  & 0.36 & 0.52 & ms/wd \\
\object{[CW83]~1735+22}       & \, 1.280(6)      & 2152.16(5)       &   20.6 & 0.4  & 104.6 & 0.5 & 0.152(3)  & 0.53 & 0.78 & wd    \\ 
\object{HD~171858}         & \, 1.63280(5)    & 2153.3684(5)     &   62.5 & 0.1  &  87.8 & 0.2 & 0.114(1)  & 0.46 & 0.67 & wd    \\     
\object{PB~7352}           & \, 3.62166(5)    & 2151.1821(5)     &  $-$2.1 & 0.3  &  60.8 & 0.3 & 0.084(1)  & 0.40 & 0.58 & ms/wd \\
\object{Ton~S~135}         & \, 4.122(8)      & 2152.49(5)       &  $-$3.7 & 1.1  &  41.4 & 1.5 & 0.030(3)  & 0.26 & 0.36 & ms/wd    \\
\object{CD$-24\degr$~731}  & \, 5.85(30)     & 2153.5(2)        &  20 &   5    &  63    & 3   & 0.15(3)  & 0.55   & 0.81 & wd    \\
\object{PHL~457}           & \, $<3$   \\
\object{LB~1516}           & \, $7\ldots27$    \\ 
\hline\\
\end{tabular}
\parbox[t]{\textwidth}{
$^*$: \begin{minipage}[t]{16cm}\object{HD~188112} is not an EHB star but a progenitor of a helium core white dwarf
with a mass of only $M=0.24~{\rm M}_{\sun}$ (Heber et al. \cite{Heb03}).
\end{minipage} \\
$^{**}$: \begin{minipage}[t]{16cm}Heber et al. (\cite{Heb03}) suggest that the compact secondary could also 
be a neutron star or a black hole.\\
\end{minipage}} 
\end{table*}
The alternative, a somewhat eccentric orbit, is more plausible.
To test this we took the best fitting sinusoidal for each star,
calculated a set of theoretical RV curves with varying eccentricities
and periastron angles $\Omega$,
and fit these curves to the observed RV values. We double checked the 
results by using an additional  program (Mayer, priv. comm.) which 
fits all parameters, including eccentricity and periastron angles,
simultaneously to the observed data points.

Both methods give exactly the same results:
The observed RV points for three stars
can be reproduced 
best assuming a non-circular orbit (see Fig. \ref{plot_ecc}).
For \object{PG~1232$-$136}
we determined an eccentricity of 
$\epsilon=0.060\pm0.005$ and a periastron angle of
$\Omega=162.5\grad \pm 0.5\grad$,
for \object{[CW83]~1419$-$09} the points match best by applying an eccentricity of
$\epsilon=0.039\pm0.005$ together with a periastron angle of
$\Omega=355.6\grad \pm 0.5\grad$,
and for \object{PG~0133+114} the observed RV points are fitted almost
perfectly by using an eccentricity of 
$\epsilon=0.025\pm0.005$  and a periastron angle of
$\Omega=101.5\grad \pm 0.5\grad$.

The data points for \object{PB~7352} and \object{PG~0001+275} can also be matched 
significantly better for non-zero eccentricities than assuming circular 
orbits (see Fig. \ref{plot_ecc}).
For \object{PB~7352} an eccentricity of 
$\epsilon=0.024\pm0.01$ and a periastron angle of 
$\Omega=85.8\grad \pm 1.0\grad$ 
results, while fore
\object{PG~0001+275} we determine an eccentricity of
$\epsilon=0.022~\pm~0.015$ together with a periastron angle of 
$\Omega=222.8\grad~\pm~2.0\grad$

As the circularization time-scale strongly depends on the period
($t_{\rm cir} \sim P^{49/12}$, Tassoul \& Tassoul \cite{Tas92})
one would expect that the eccentricity would be correlated
with the period.
However,
for our five stars this is not the case; i. e. the star with the 
longest period (\object{PB~7352}) does not have the largest eccentricity, nor does the 
star with the shortest period (\object{PG~1232$-$136}) have the smallest eccentricity.
However, lacking a better explanation, we suggest that their orbits
are probably not circular, making these close binaries the first 
for which eccentric orbits are detected.  
\subsection{Mass determination of the unseen companions}
\label{mass_determination}
Since the stars are single-lined binaries, we can only derive the mass function
\begin{eqnarray*}
\qquad f_m = \frac{M_{\rm comp.}^3 \sin^3(i)}{(M_{\rm sdB}+M_{\rm comp.})^2} =
       \frac{P K^3}{2 \pi G}.
\end{eqnarray*}
\label{mass_function}
Using the mass function and adopting the canonical mass of the sdB
star of $M=0.5~{\rm M}_{\sun}$ (Heber \cite{Heb86}), lower limits 
(inclination $i=90^\circ$) to the masses 
of the unseen companions can be derived.
From a statistical point of view, most probable masses can also be 
calculated by adopting an average inclination 
of $i=52\degr$ for the invisible secondaries.

For the five systems for which we detected a somewhat non-circular orbit, 
the mass determination stays the same assuming a circular orbit, because the 
discovered very small eccentricities do not alter the results.
Also for the four stars for that alias periods exist which cannot be ruled out
(cf. Section \ref{period_determination}),
the results do not change if the second (or third) best period is applied.
As described in Sect. \ref{period_determination} for 
two stars
(\object{PHL~457} and \object{LB~1516})
no meaningful periods could be determined due to the low numbers
of data points. Therefore we 
shall not discuss these two stars further.

The mass functions and the lower limits to the companion masses, as well as their
most probable values, are denoted in Table \ref{orbital_parameters}.
\subsection{Comparison with previous results}
Two of our program stars, \object{HD~171858} and 
\object{PG~0133+114}, have been
discovered independently to be RV variable by 
Morales-Rueda et al. (\cite{Mor03}). 

For \object{HD~171858}, they determined 
orbital parameters which differ noticeably from our values
($\Delta P\approx 0.1$~days, $\Delta K\approx 6$~\kms, and
$\Delta \gamma_0\approx 11$~\kms).
However, due to the poor phase coverage of Morales-Rueda  et al.
(only 12 RV measurements all close to the maximum and minimum
of the RV curve in contrast to our 48 RV data points with an
almost perfect sampling), the differences are understandable.
Their estimate of the minimum mass for the invisible companion of
$0.51 {\rm M}_{\sun}$ is in reasonable agreement with our result
of $0.46 {\rm M}_{\sun}$.\footnote{Morales-Rueda et al. (\cite{Mor03}) used 
for the mass of the sdB primary, as we do, the canonical mass of $M=0.5 {\rm M}_{\sun}$.}

For \object{PG~0133+114} Morales-Rueda et al. used 18 RV measurements to
determine a period of $P=1.2382(2)$ days 
(number in parentheses give the uncertainty of the
last given digit) which is only
slightly different to our result of $P=1.23787(3)$ days (20 RV measurements).
Their RV semi amplitude agrees very 
well 
with our value 
($\Delta K\approx 1$~\kms).
Only their system velocity of $\gamma_0 = +6 \pm 1$~\kms\ 
%
does not match with our result of  
$\gamma_0 = -0.3 \pm 0.2$ \kms.
The resulting lower limit of $M=0.388 {\rm M}_{\sun}$, 
determined by Morales-Rueda et al. 
for the companion, however,  
is again close to our result of $0.36~{\rm M}_{\sun}$.

New measurements for \object{HD~188112} allowed us to increase the accuracy of the
orbital parameters. 
Our results of $P=0.6065812(5)$~days, $K = 188.4 \pm 0.2$~\kms, and
$\gamma_0 = +26.7 \pm 0.2$~\kms, are  consistent with 
our former determinations
($P=0.606585(2)$~days, $K = 188.3 \pm 0.5$~\kms, and
$\gamma_0 = +26.6 \pm 0.3$~\kms, Heber et al. \cite{Heb03}).
%
%
%
%
%
%
%
%
%
%
%
\subsection{Nature of the unseen companions}
The companions are most likely either late type main sequence stars
or white dwarfs.
Spectral signatures of degenerate companions would not be
detectable at optical wavelength due to the faintness of the white dwarfs.  
However, spectral features arising from a late type main sequence star
could be detectable if the star is sufficiently bright.
Therefore we have searched for absorption
lines which are prominent in cool stars, e.g. the \ion{Ca}{ii} H and K
line, the G-band, the \ion{Mg}{i} triplet at 5167~\AA, 5173~\AA, and 5184~\AA,
or the \ion{Ca}{ii} triplet (CaT) at 8498~\AA, 8542~\AA, and 8662~\AA\ (cf. 
Jeffery \& Pollacco \cite{Jef98}). None are found. A more detailed 
analysis will be presented in the forthcoming Paper II.

%
%
%
%
%

We estimate that any cool main sequence star that contributes more than
10\% of light in the $I$ band should be detectable via the CaT search.
Even at a spectral type as late as M1 the  CaT lines are strong with
equivalent width larger than 1~\AA\ (Jones, Alloin \& Jones, \cite{Jon84}).
Adopting $M(I)=4.6$~mag for the sdB star
this corresponds to a companion spectral type of $\approx$ M1 or a mass of
$\approx 0.45~{\rm M}_{\sun}$ (Drilling \& Landolt, \cite{Dri00}).
Accordingly we classify the companion as a white dwarf if its minimum
mass exceeds $0.45~{\rm M}_{\sun}$. 
This is the case for \object{CD$-24\degr$~731},
\object{[CW83]~1735+22}, \object{HD~171858}, and \object{HD~188112}.
If we adopt the statistically most likely inclination angle, the
corresponding minimum companion mass of 
\object{PB~7352}, \object{PG~0133+114}, \object{PG~1232$-$136}, \object{Ton~S~183}, 
and \object{[CW83]~1419$-$09}
exceeds $0.45~{\rm M}_{\sun}$.
Therefore it is likely that several of those systems host a white dwarf.
%
In all other cases the companion type can not
be constrained further from our observations, they could be either main
sequence stars or white dwarfs\footnote{The mass distribution of
white dwarfs peaks near $0.6 {\rm M}_{\sun}$ (see e.g. Liebert, Bergeron \& Holberg, 
\cite{Lie05}, Madej, Nale\.zyty \& Althaus, \cite{Mad04}, and references therein). 
However, white dwarfs with masses as low as 
$0.2 {\rm M}_{\sun}$ do exist (see e.g. Heber et al. \cite{Heb03},
Liebert et al. \cite{Lie04}).}. 
The minimum companion masses for \object{BPS~CS~22169$-$0001} and \object{CPD$-64\degr$~481}
are so small ($0.03~{\rm M}_{\sun}$ and $0.05~{\rm M}_{\sun}$, respectively) 
that they may be brown dwarfs if the inclination is larger than 
$20\degr$ and $38\degr$, respectively.
 
Table \ref{orbital_parameters} summarizes the probable nature for all 
companions to our program stars.  
\section{Discussion}
\label{discussion}
The unseen companions 
of three stars are white dwarfs whereas in the other cases they 
are either low mass main 
sequence stars or white dwarfs. The periods are in almost all cases
shorter than ten days (dictated by our search strategy). 
This indicates that all observed
RV variable sdB stars have evolved through at least one common envelope  
phase, consistent with the theoretical prediction of Han et al. (\cite{Han03}).

After such a common envelope phase the orbit of the resulting
close binary system should be 
circular, irrespective of
a possible former eccentric orbit, like found for  
all former observed close hot subdwarf or white dwarf binary systems.
Also from our analysis, most of our radial velocity curves can 
be reproduced best by assuming a circular orbit.
However, for five systems 
(which comprise one third of our sample)
we detected periodic deviations from fitted sinusoidal curves. 
%
%

These deviations can be removed almost perfectly 
for \object{PG~1232$-$136}, \object{[CW83]~1419$-$09}, and \object{PG~0133+114} 
by introducing small
eccentricities of $\epsilon=0.025$-0.06.
Also for \object{PB~7352} and \object{PG~0001+275} the observed data points can be
matched significantly  better assuming a very small eccentricity
of $\epsilon\approx0.02$.
%
%
%
%
%
%

Unfortunately, for all five stars it is unclear whether their companions 
are main sequence stars or white dwarfs, i.e. we do not know whether
the stars have evolved during one or two common envelope phases.
If their companions are white dwarfs, it is really hard to believe that
the orbits of the stars remain eccentric although the systems
have undergone two common envelope phases.
On the other hand, if the companions are main sequence stars,
the observed very small eccentricities could maybe be remnants of a former 
highly eccentric orbit. 

Could a small remaining eccentricity be a tracer of main sequence companions 
(only one mass exchange phase)?
Further investigations are necessary to verify this assumption. 
In particular, it would be most rewarding to measure light curves and search 
for a reflection effect indicative of a main sequence companion.
%
%
\section{Summary and conclusion}
We have determined the radial velocity curves for 17 bright binary subdwarf B 
systems using high precision radial velocity measurements from 
high S/N optical high-resolution spectra,
and derived orbital parameters for the 15 unambiguous cases.
The companions are unseen in the spectra. 

For most systems the orbits are circular (eccentricity $\epsilon<0.02$,
i.e. our detection limit).
However, for five sdB stars 
we discovered that their orbits are probably non-circular
with small eccentricities of $\epsilon=0.022$--0.060.
These close binaries are the first for which 
eccentric orbits have been detected.

Using the canonical mass for the sdB primary of $M=0.5~{\rm M}_{\sun}$
and the mass function, the 
nature of
the invisible secondaries for all program stars could be constrained. 
Three systems 
consist of a sdB star and a white dwarf
because the companion mass exceeds $\sim 0.45~{\rm M}_{\sun}$.
The companions of two systems 
are possibly brown dwarfs.
For all other systems 
the nature of the unseen companions
remain unclear; they could either be main sequence stars or
white dwarfs.



Important questions remain to be answered. 
Are the periodic deviations really due to eccentric orbits?
If so, are the radial velocity curves of all other close binary sdB systems, which have 
been determined by other groups really consistent with circular orbits or 
were the observations performed so far simply too inaccurate
to detect such small eccentricities?
To verify or clarify the nature of the invisible companions
more high-precision measurements of  radial velocity variable
sdB systems and further investigations
such as searching for eclipses or reflection effects
are necessary.

The next step will
be to determine the binary frequency of our sample.
Additional observations, however, are required to achieve this goal.
Forthcoming papers will also present the results of quantitative 
spectral analyses to measure atmospheric parameters and abundances,
and the kinematics of the sample population will be studied to assign 
their membership. 
%
%
\begin{acknowledgements}
The authors appreciate the help, support, and valuable assistance
provided by the staff of the Calar Alto observatory, Spain, 
and the ESO La Silla observatory, Chile, during our visits or during
service-mode observations.
Many thanks goes also to Horst Drechsel and Ralf Napiwotzki for fruitful discussions
and to Pavel Mayer for providing us with a program code
to derive orbital parameters for stars with eccentric orbits.
H.~E. and C.~K. acknowledge financial support by the German 
research foundation DFG under grants He~1354/30$-$1, Na~365/2$-$2, and
He~1356/40$-3$, and for several travel grants to the Calar Alto observatory.
M.~A. is supported by the FONDAP 1501 0003 Centre for Astrophysics.
We made extensive use of NASAs Astrophysics Data System Abstract Service (ADS),
the SIMBAD database, operated at CDS, Strasbourg, France, 
and the Digital Sky Survey (DSS).
\end{acknowledgements}

\end{document}